\documentstyle[preprint,aps]{revtex}

\def\Journal#1#2#3#4{{#1}{\bf #2},#3 (#4)}
\def\prd{{\em Phys. Rev.} D}
\def\pra{{\em Phys. Rev.} A}
\def\prb{{\em Phys. Rev.} B}
\def\npb{{\em Nucl. Phys.} B}
\def\prl{\em Phys. Rev. Lett.}
\def\cmp{\em Comm. Math. Phys.}

\def\plb{{\em Phys. Lett} B}

\newcommand{\beq}[1]{\begin{equation}\label{#1}}
\newcommand\eeq{\end{equation}}
\newcommand{\ba}[1]{\begin{eqnarray}\label{#1}}
\newcommand{\baa}{\begin{eqnarray}}
\newcommand\ea{\end{eqnarray}}
\newcommand{\bee}{\begin{equation}}
\def\nn{\nonumber \\}
\def\l{\lambda}

\newcommand{\h}{Hamiltonian}

\def\hlf{\frac{1}{2}}

\begin{document}
\title{Duality and quasiparticles in the Calogero-Sutherland model:
 Some exact results}
\author{Ivan Andri\'c and Larisa Jonke
\footnote{e-mail address: 
andric@thphys.irb.hr \\ \hspace*{2.2cm} 
larisa@thphys.irb.hr }}
\address{Theoretical Physics Division,\\
Rudjer Bo\v skovi\'c Institute, P.O. Box 180,\\
HR-10002 Zagreb, CROATIA}
\maketitle
\begin{abstract}
\widetext
The quantum-mechanical many-body system with the potential proportional to 
the pairwise inverse-square  distance possesses a strong-weak coupling 
duality. Based on this duality, particle and/or quasiparticle states are 
described as SU(1,1) coherent states. The constructed quasiparticle states are
of hierarchical nature.
\narrowtext
\end{abstract}
PACS numbers: 03.65.Nk 05.30.Pr 05.45.Yv
\section{Introduction}

The one-dimensional pairwise inverse-square distance interaction of $N$ 
particles, and their generalizations under the name of
Calogero-Sutherland-Moser models (CSMM)\cite{c,s,m} have so far appeared in a 
variety of different physical contexts. They are related to the random 
matrix model\cite{meh} and the two-dimensional Yang-Mills theory\cite{min}.
They also represent an example of generalized exclusion statistics\cite{hal}, 
and quantum spin chains with long-range interactions\cite{hall}. 
CSMM's describe edge states in the quantum Hall system\cite{Ka} and the 
Chern-Simons theory\cite{prd}.

We still lack a local 
canonical field-theoretical formulation of CSMM's\cite{Am}, 
but a collective-field theory\cite{Je} can be established\cite{An,Sen} in 
the large-$N$ limit, and it connects CSMM's with $2$d gravity\cite{dj2d}. 
A deeper understanding of the models can be gained by 
exploring various solutions. CSMM's are exactly solvable and integrable, 
both classically and quantum-mechanically. From the quantum Lax 
formulation\cite{Bar}
 we can find infinitely many commuting conserved operators, and 
the underlying algebraic structure should reveal the large degeneracy 
structure of CSMM's. Its eigenfunctions are known to be symmetric 
polynomials\cite{Sta,Gur} but, to these days, the only explicit 
form have been the original wave functions found by Calogero\cite{c}.
In the collective-field formalism it has been 
 found that there exist static solitons 
in the Bogomol'nyi limit and moving solitons as solutions of the equations of 
motion\cite{Jev,Po,Sen}.

We would like to show that a specific CSMM on the line (called the Calogero-
Moser model) with the \h \ ($\hbar=m=1$) given by
\beq 1
H_{CM}=\hlf\sum_{i=1}^Np_i^2+\frac{\l(\l-1)}{2}\sum_{i\neq j}^N\frac{1}
{(x_i-x_j)^2}\eeq
admits a family of new solutions describing the quasiparticles. 
We shall construct the dynamics of quasiparticle
 states using the subgroup SU(1,1)\cite{Pere} of 
the $W_{\infty}$ algebra\cite{Bar}. It has been conjectured\cite{Gura} 
that Calogero 
scattering eigenfunctions are generalized coherent states related to the 
representations of SU(1,1). A proper treatment of the 
center-of-mass degrees of freedom is important in order to preserve the 
translational invariance of the model. Using the weak-strong coupling 
duality\cite{Gau} between particles and quasiparticles, we can extend the 
coherent-state representation to the case with both particles and 
quasiparticles.

\section{SU(1,1) algebra}

Let us first construct the representation of a spectrum 
generating algebra for the \h \ (\ref{1}) as a generator. In fact, there is a 
larger symmetry group \cite{Bar}, but here we deal only with 
quasiparticles related to the subgroup $SU(1,1)$.  
It is  convenient to extract the Jastrow factor from the wave function and 
perform  a similarity transformation of the \h \ into
\beq 3
\prod_{i<j}^N(x_i-x_j)^{-\l}(-H_{CM})\prod_{i<j}^N(x_i-x_j)^{\l}=
\hlf\sum_{i=1}^N\partial_i^2+\frac{\l}{2}\sum_{i\neq j}^N\frac{1}{x_i-x_j}
(\partial_i-\partial_j) .\eeq
Owing to the translational invariance of the model we should introduce 
completely translationally invariant variables
\bee
\xi_i=x_i-X,\;\;\partial_{\xi_i}\xi_j=\delta_{ij}-\frac{1}{N}.\eeq
Here we have introduced the center-of-mass coordinate 
$X=\frac{1}{N}\sum_{i=1}^Nx_i$ and its canonical conjugate 
$\partial_X=\sum_{i=1}^N\partial_i=iP_{\rm tot}$. As an $SU(1,1)$ generator 
we take \h \ with eliminated center-of-mass degrees of freedom:
\beq 2 
T_+(x,\l)=\hlf\sum_{i=1}^N\left(\partial_i-\frac{1}{N}\partial_X\right)^2
+\frac{\l}{2}\sum_{i\neq j}^N\frac{1}{x_i-x_j}
(\partial_i-\partial_j)=\hlf\sum_{i=1}^N\partial_{\xi_i}^2+\frac{\l}{2}
\sum_{i\neq j}^N\frac{1}{\xi_i-\xi_j}(\partial_{\xi_i}-\partial_{\xi_j}).\eeq
Owing to the scale and special  invariance we introduce two 
additional generators:
\ba 4 
T_0(x,\l)&=&-\hlf(\sum_{i=1}^Nx_i\partial_i-X\partial_X+E_0-\frac{1}{2})=
-\hlf(\sum_{i=1}^N\xi_i\partial_{\xi_i}+E_0-\hlf),\nn
T_-(x,\l)&=&\hlf\sum_{i=1}^Nx_i^2-\frac{N}{2}X^2=\frac{1}{4N}\sum_{i\neq j}^N
(x_i-x_j)^2=\hlf\sum_{i=1}^N\xi_i^2, \ea
and, after performing  some calculation, we can verify
\beq x
[T_+,T_-]=-2T_0,\;[T_0,T_{\pm}]=\pm T_{\pm}.\eeq
This  is the usual SU(1,1) conformal 
algebra\cite{Bar,Gura}, with the Casimir operator
$\hat C=T_+T_--T_0(T_0-1)$.
In the definition (\ref{4}) of the operator $T_0$
the  constant $E_0$
is $E_0=\frac{\l}{2}N(N-1)+\frac{N}{2}$ for consistency reasons,
 and $-1/2$ appears
after removing the center-of-mass degrees of freedom.

After having established the representation of $SU(1,1)$ algebra, we show that 
the Calogero solutions are 
completely determined assuming the zero-energy solutions are known:
\baa\label{21}
&&T_+(x,\l)P_m(x_1,x_2,...,x_N)=0,\nn
&&T_0(x,\l)P_m(x_1,x_2,...,x_N)=\mu_mP_m(x_1,x_2,...,x_N),\;
\mu_m=-\hlf(m+E_0-\hlf).\ea
Calogero has proved that the functions $P_m(x_1,x_2,...,x_N)$ are scale- and
translationally invariant homogeneous multivariable polynomials of 
degree $m$, written in the center-of-mass frame. There are no general explicit 
representations of these polynomials, except in the case including 
quasiparticles, which we derive below.
Let us assume that a nonzero eigenstate of the operator $T_+(x,\l)$ 
is  of the general form
\baa\label{22}
&&\Psi(T_-,T_0,T_+)P_m(x_1,x_2,...,x_N)=\sum_{p,q,n}c_{pqn}T_-^pT_0^qT_+^n
P_m(x_1,x_2,...,x_N)\nn &&=\Psi(T_-,T_0)P_m(x_1,x_2,...,x_N)
=\Psi_m(T_-)P_m(x_1,x_2,...,x_N).\ea
Using (\ref{x}) we can derive the formula
\bee\label{23}
[T_+,f(T_-)]=T_-f''(T_-)-2f'(T_-)T_0.\eeq
Using (\ref{x}) and  the eigenvalue equation
\bee
-T_+\Psi_m(T_-)P_m(x_1,x_2,...,x_N)=E\Psi_m(T_-)P_m(x_1,x_2,...,x_N)\eeq
we obtain the Calogero solution (in his notation $p=\sqrt{2E}$, $r^2=2T_-$):
\baa\label{sol}
\Psi_m(T_-)P_m
&\sim& T_-^{(1-m-E_0+1/2)/2}{\cal Z}
_{m+E_0-3/2}(2\sqrt{ET_-})P_m(x_1,x_2,...,x_N)\nn
&\sim&r^{-(m+E_0-3/2)}{\cal Z}_{m+E_0-3/2}(pr)P_m(x_1,x_2,...,x_N).\ea
We can also show that already at the first level the Calogero \h \ (\ref{1})
for symmetric wave function 
possesses a quasiparticle solution. 
For $P_{\kappa}(\xi_i)=\prod_i^N\xi_i^{\kappa}$, with $\kappa=1-\l-1/N$, 
the equation
\bee\label{12nova}
T_+\prod_{i=1}^N\xi_i^{\kappa}\Psi_(T_-)=-E\prod_{i=1}^N\xi_i^{\kappa}\Psi_(T_-)\eeq has a Bessel function quasihole solution sitting at the origin of 
the Calogero system. Up to the $1/N$ correction for $\l$, this is the 
type of solution already found in the collective-field approach\cite{Jev}.

\section{Duality}

The Calogero-Sutherland model is a rare quantum-mechanical system where 
a strong-weak coupling duality exists
\cite{Gau,Sta} relating various physical quantities for the constants of 
interaction $\l$ and $1/\l$. Depending on the parameters, the  duality 
exchanges particles with quasiparticles. We  show how to find a 
solution of the model with quasiparticles using duality relations.
From the collective-field-theory approach to the problem we know that a wave 
function describing the holes (or lumps) has a prefactor of the form
\beq b V^{\kappa}(x-z)=\prod_{i,\alpha =1}^{N,M}(x_i-Z_{\alpha})^{\kappa}
,\; \alpha=1,...M,\eeq
where $Z_{\alpha}$ denotes $M$ zeros of the wave function, describing 
positions of $M$ quasiparticles.
The duality is displayed by the following relations:
\ba c
P_{\rm tot}(x)V^{\kappa}(x-z)&=&-P_{\rm tot}(z)V^{\kappa}(x-z),\nn
T_0(x,\l)V^{\kappa}(x-z)&=&\left\{-T_0(z,\frac{\kappa^2}{\l})
-\frac{1}{2}[\kappa NM+
\epsilon_0(N,\l)+\epsilon_0(M,\frac{\kappa^2}{\l})]\right\}V^{\kappa}(x-z),\nn
T_+(x,\l)V^{\kappa}(x-z)&=&\left\{-\frac{\l}{\kappa} T_+(z,\frac{\kappa^2}{\l})+
\frac{1+\l/\kappa}{2}\sum_{i,\alpha}^{N,M}\frac{\kappa(\kappa-1)}{(x_i-
Z_{\alpha})^2}\right\}V^{\kappa}(x-z),\ea 
where the operator $T_{0,\pm}(z,\frac{\kappa^2}{\l})$ denotes an operator with
the same functional dependence on $Z_{\alpha}$ as that of the 
operator $T_{0,\pm}(x,\l)$
on $x_i$,
 with the coupling constant $\l$ changed into $\kappa^2/\l$.
In addition to these duality relations we  take that the center-of mass 
coordinates are identical, namely, 
$X=Z$, ($Z=\frac{1}{M}\sum_{\alpha=1}^MZ_{\alpha}$) at the end of 
calculations.
The duality also places an interesting restriction on the number of 
quasiparticles $M$. 
Namely, the number of quasiparticles is dermined by the coupling constant
 $\l$ and is proportional to the number of particles $N$:
\bee\label{rel} M=-\frac{\l N}{\kappa}.\eeq
Here we have manifest duality: if we interchange particles and quasiparticles 
($N\leftrightarrow M$), then $\l$ goes to $\kappa^2/\l$. For 
$\kappa=1$, the relation (\ref{rel}) was conjectured in Ref.\cite{Gau}.
Let us introduce new collective generators for the 
system of particles and quasiparticles:
\ba d
{\cal T}_+&=&T_+(x,\l)+\frac{\l}{\kappa} T_+(z,\frac{\kappa^2}{\l})-
\frac{(\l+\kappa)(\kappa-1)}{2}\sum_{i,\alpha}^{N,M}\frac{1}{(x_i-Z_{\alpha})^2
},\nn
{\cal T}_0&=&T_0(x,\l)+T_0(z,\frac{\kappa^2}{\l}),\nn
{\cal T}_-&=&T_-(x,\l)+\frac{\kappa}{\l}T_-(z,\frac{\kappa^2}{\l}).\ea
It can be easily checked that the above generators satify the 
$SU(1,1)$ conformal 
algebra. In terms of the generators (\ref{d}), the duality relation 
(\ref{c}) turns 
out to be a sufficient condition for solving the Calogero model with 
quasiparticles:
\ba s
{\cal T}_+V^{\kappa}(x-z)&=&0,\nn
{\cal T}_0V^{\kappa}(x-z)&=&-\frac{(N+M)(\kappa+1)-2}{4}V^{\kappa}(x-z).\ea
We interpret the  operator ${\cal T}_+$ as a \h \ for a more general problem.
After performing a similarity transformation of ${\cal T}_+$, we obtain the 
master \h \ for particles and quasiparticles,
\baa\label{z}
H(x,z)&=&-\hlf\sum_{i=1}^N\partial_i^2+\frac{\l(\l-1)}{2}\sum_{i\neq j}^N
\frac{1}{(x_i-x_j)^2}\nn
&+&\frac{\l}{2\kappa}\sum_{\alpha=1}^M\partial_{\alpha}^2+
\frac{\kappa^2}{2\l}\left(\frac{\kappa^2}{\l}-1\right)\sum_{\alpha\neq\beta}^M
\frac{1}{(Z_{\alpha}-Z_{\beta})^2}-\hlf\left(1+\frac{\l}{\kappa}\right)
\sum_{i,\alpha}^{N,M}\frac{\kappa(\kappa-1)}{(x_i-Z_{\alpha})^2}.\ea
We see
that the quasiparticle mass is $\kappa/\l$, and the factor in front of the 
interaction term is $\kappa(\kappa -1)/2$ times the inverse reduced mass of the 
particles and quasiparticles (remember that we set the particle mass to one). 
This is the second hierarchical level \h \ with the solution for the 
given energy $E=k^2/2$
\bee\label{solhier}
\Psi(x,z;k)=\prod_{i<j}^N(x_i-x_j)^{\l}\prod_{\alpha<\beta}^M(Z_{\alpha}-Z_{\beta})^{\kappa^2/\l}(kR)^{-b}{\cal Z}_b(kR)\prod_{i,\alpha}^{N,M}(x_i-Z_{\alpha})^{\kappa},\eeq
where the index of the Bessel function is $b=\kappa MN+E_0(N,\l)+E_0(M,\kappa^2/\l)-2$. The solution has been found following the procedure outlined in 
Eqs.(\ref{22}-\ref{sol}). To get more insight into the hierarchical \h \
we can eliminate quasiparticle degrees of freedom by performing the 
appropriate derivatives in ${\cal T}_+$ and putting $Z_{\alpha}=0$. 
The remaining equation is the Calogero first hierarchical level 
for quasiparticles situated at the origin (\ref{12nova}).

\section{Coherent States}

Let us show that the Calogero solution (\ref{sol}) let in the operator form 
is a generalized Barut-Girardello coherent 
state of the $SU(1,1)$ group \cite{Pere}. In this case the coherent state 
is defined as an eigenvector of the "annihilation" operator $T_+(x,\l)$.
The Calogero solution (\ref{sol}) can be recast in the form of a 
coherent state by specifying the coefficients $c_{pq0}$:
\bee
\Psi(T_-,T_0)=\sum_{p=0}^{\infty}\frac{E^p}{p!}T_-^p\prod_{i=1}^pF(T_0-i+1)
=\exp(ET_-F(T_0)).\eeq
The operator $A^+\equiv T_-F(T_0)$ plays the role of a "creation" 
operator and the function $F(T_0)$ can be determined such that  $A^+$ 
is the canonical conjugate of $T_+(x,\l)$, i. e., $[T_+,A^+]=1$.
Using the $SU(1,1)$ algebra and demanding that the canonical commutation 
relation is valid for the zero energy solutions, we obtain
\bee\label{A+}
A^+=T_-\frac{-T_0+1-\frac{m+E_0-1/2}{2}}{\hat C+T_0(T_0-1)}.\eeq
Now, the translationally invariant Calogero eigenfunctions can be written as
\ba a \Phi(x_1,x_2,...,x_N)&=&
\exp(\frac{k^2}{2}A^+)P_m(x_1,x_2,...,x_N)\nn
&\sim&
(kr)^{-b}{\cal Z}_b(kR)P_m(x_1,x_2,...,x_N),\ea
where ${\cal Z}_b$ is a Bessel function, $b=m+E_0-\frac{3}{2}$, and $r^2=2T_-$.

Finnaly, following the steps outlined above, 
we construct the canonically conjugate operator ${\cal A}^+$
\bee
{\cal A}^+={\cal T_-}\frac{-{\cal T}_0+2-\frac{(N+M)}{2}(\kappa+1)}
{\hat{\cal C}+{\cal T}_0({\cal T}_0-1)}, \eeq
and then a coherent state of particles and quasiparticles follows as
\bee
\Psi(x,z;k)=\prod_{i<j}^N(x_i-x_j)^{\l}\prod_{\alpha<\beta}^M(Z_{\alpha}-Z_
{\beta})^{\kappa^2/\l}
\exp(\frac{k^2}{2}{\cal A}^+(x,z))V^{\kappa}(x-z).\eeq
Applying the operator ${\cal A}^+$ to $V^{\kappa}$ we obtain the solution 
(\ref{solhier}).

\section{Conclusion}

In summary, we have constructed the master \h \ for particles and their dual 
quasiparticles. The solutions of this \h \ in the operator form has been 
found to be generalized $SU(1,1)$ coherent states. Formally, this is also a 
solution to the Calogero two-family problem\cite{c} with different masses and 
coupling  constants. Owing to duality, the construction of the master \h \ 
and its solutions are of hierarchical nature\cite{hier}. In fact, each new 
family represents a new hierarchical level and is obtained by introducing  a 
new prefactor (in the wave function) and extending the $SU(1,1)$ generators 
by corresponding terms for quasiparticles. For example, the third family 
introduced by $V^{\gamma}(z-y)$, where $y$ represents positions of new 
quasiparticles, will have the mass $\kappa/\gamma$ and the statistical 
factor $\frac{\gamma^2}{\kappa^2\l}$. There are possible more complicated 
constructions owing to the fact that higher dynamical groups then $SU(1,1)$ 
exists for the Calogero model. We may observe that in many respects 
(duality, hierarchy, statistics) the CSMM very much resembles $2$-dimensional 
quasiparticles appearing in the fractional quantum Hall effect. 
An interestiagn open question is how to formulate this hierarchy in field 
theory context.

\section*{ Acknowledgment}

This work was supported by the Ministry of Science and Technology of the
 Republic of Croatia under contract No. 00980103.


\end{document}